\documentclass[lettersize,journal]{IEEEtran}
\usepackage{amsmath,amsfonts}
\usepackage{algorithmic}
\usepackage{array}
\usepackage[caption=false,font=normalsize,labelfont=sf,textfont=sf]{subfig}
\usepackage{textcomp}
\usepackage{stfloats}
\usepackage{url}
\usepackage{verbatim}
\usepackage{graphicx}

\usepackage{algorithm}
\usepackage{algorithmic}
\usepackage{amsmath, amssymb}
\usepackage{xcolor}
\definecolor{citecolor}{HTML}{0071bc}
\usepackage[colorlinks, linkcolor=red,  anchorcolor=blue, citecolor=citecolor]{hyperref} 

\usepackage{xcolor}
\definecolor{SeaGreen4}{RGB}{0,205,102} 
\definecolor{SlateBlue}{RGB}{106,90,205} 
\definecolor{DarkRed}{RGB}{178,34,34} 
\usepackage{booktabs}
\usepackage{pifont}

\newcommand{\cmark}{\ding{51}}
\newcommand{\xmark}{\ding{55}}

\def\BibTeX{{\rm B\kern-.05em{\sc i\kern-.025em b}\kern-.08em
    T\kern-.1667em\lower.7ex\hbox{E}\kern-.125emX}}
\usepackage{balance}
\begin{document}
\title{SafeDivertor: Faithful Divertor Heat Flux Reconstruction from Macroscopic Plasma State Signals via Time-Frequency Prior Exploitation}
\author{Hao Si, Zehua Chen*, Qingquan Yang*, Xiao Wang, Dengdi Sun, Wanli Lyu, Gaoting Chen, \\ Guosheng Xu, Hang Su, Jin Tang, and Jun Zhu
\thanks{$\bullet$ Hao Si, Xiao Wang, Wanli Lyu, and Jin Tang are with the School of Computer Science and Technology, Anhui University, Hefei 230601, China. (email: e24201034@stu.ahu.edu.cn, \{xiaowang, lwl, tangjin\}@ahu.edu.cn)}
\thanks{$\bullet$ Zehua Chen, Hang Su, and Jun Zhu are with the Department of Computer Science and Technology, Tsinghua University, Beijing, China. (email: \{zhc23thuml, suhangss, dcszj\}@tsinghua.edu.cn)}
\thanks{$\bullet$ Dengdi Sun is with the School of Artificial Intelligence, Anhui University, Hefei 230601, China. (email: sundengdi@163.com)}
\thanks{$\bullet$ Qingquan Yang, Gaoting Chen, and Guosheng Xu are with the Institute of Plasma Physics, Chinese Academy of Sciences, Hefei 230601, China. (email: \{yangqq, gaoting.chen, gsxu\}@ipp.ac.cn)} 
\thanks{\emph{Corresponding authors: Zehua Chen $\&$ Qingquan Yang}}
}

\markboth{IEEE TRANSACTIONS ON ***~2026}%
{How to Use the IEEEtran \LaTeX \ Templates}

\maketitle

\begin{abstract}
Divertor heat-flux analysis is essential for understanding plasma-wall interactions and protecting plasma-facing components in magnetic-confinement fusion devices, while conventional infrared-based inversion is usually performed after discharge and requires heat-conduction modeling with device-specific material properties, divertor geometry, and boundary conditions. Rather than accelerating this conventional infrared-based inversion paradigm, we introduce a new online-oriented signal-based reconstruction paradigm that directly reconstructs time-resolved radial heat-flux profiles from multi-source macroscopic plasma-state signals available during discharge. To enable systematic study of this task, we construct \textbf{DivMPS2HF}, a multi-source discharge dataset that provides the data foundation and benchmark for signal-based divertor heat-flux reconstruction. We further propose \textbf{SafeDivertor}, a task-driven framework designed to address the key challenges of signal-based heat-flux reconstruction. It employs physical prior-aware initialization to provide radial-distribution guidance for target channels, input perturbation to reduce over-reliance on specific heterogeneous signals, spectral-aware reconstruction optimization to exploit time-frequency priors and preserve transient dynamics, and progressive training to stabilize the optimization of these complementary objectives. Experiments on DivMPS2HF demonstrate that SafeDivertor achieves the best overall performance among the evaluated time-series baselines across all five metrics, establishing a new performance benchmark for signal-based divertor heat-flux reconstruction. 
The source code will be released on \url{https://github.com/Event-AHU/OpenFusion} 
\end{abstract}


\section{Introduction}

Divertor heat-flux analysis is essential for safe and high-performance fusion operation, as concentrated and transient heat loads can damage plasma-facing components~\cite{yang2020development}. 
This challenge is particularly critical for future burning-plasma devices such as ITER, where divertor heat loads are subject to strict material limits~\cite{pitts2019physics}. 
Accurate time-resolved estimation is therefore important for heat-load assessment, component protection, and future online control~\cite{aymerich2023physics}.

\begin{figure*}[!htbp]
\centering
\includegraphics[width=1.0\linewidth]{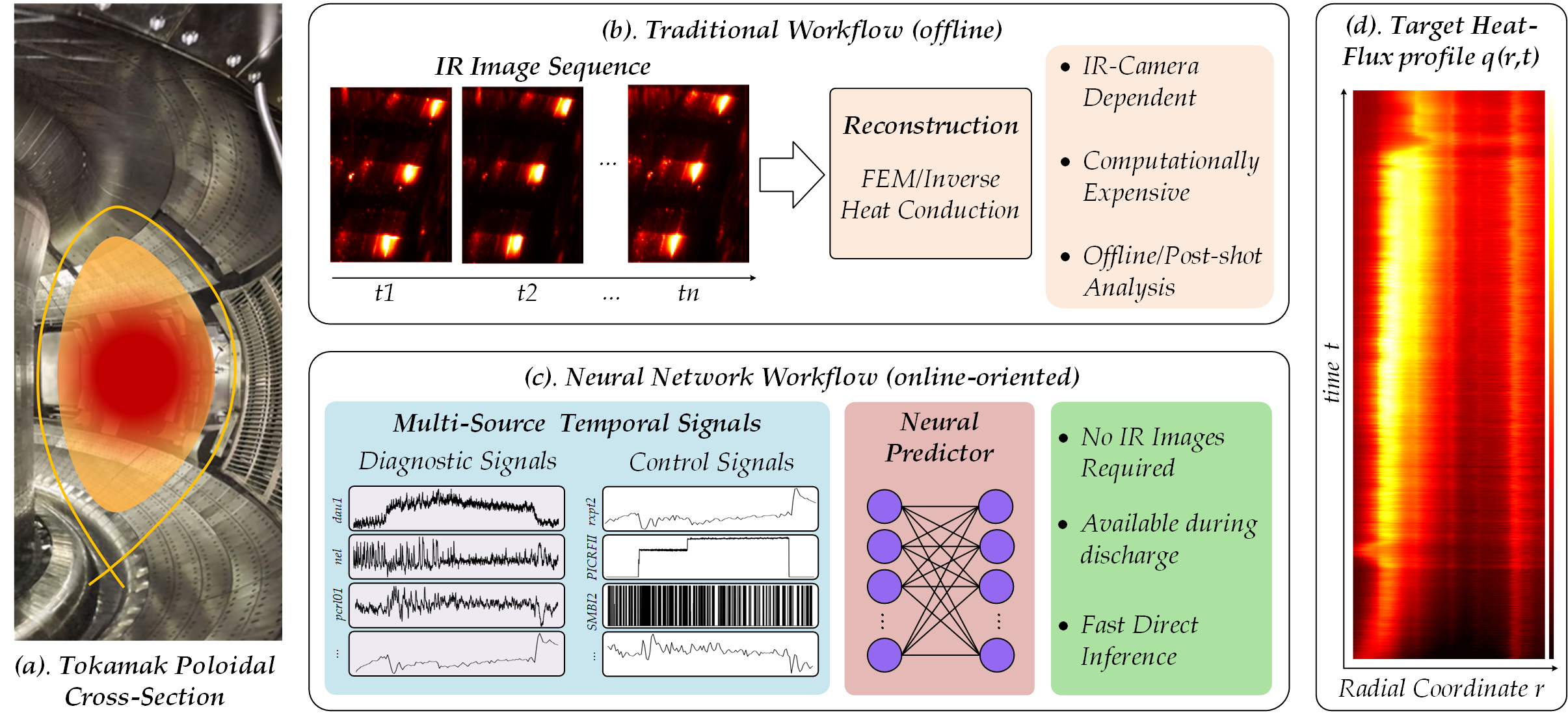}
\caption{Illustration of the conventional IR-based heat-flux inversion workflow and the proposed signal-based neural reconstruction workflow. 
(a) Tokamak poloidal cross-section. 
(b) Offline workflow based on IR image sequences and heat-conduction inversion. 
(c) Proposed online-oriented workflow that reconstructs heat-flux profiles from multi-source plasma-state signals. 
(d) Target divertor heat-flux profile \(q(r,t)\).}
\label{fig:firstimg}
\end{figure*}

As shown in Figure~\ref{fig:firstimg}(b), conventional divertor heat-flux analysis mainly follows a conventional infrared-based inversion paradigm, where the target-surface temperature is first measured and the heat-flux profile is then reconstructed using computationally expensive numerical solvers, typically for offline post-shot analysis~\cite{herrmann1995energy,lee2017thermographic,shi2017heat,yang2020development}.
Recent learning-based methods, such as Physics-Informed Neural Networks (PINNs) for accelerating heat-equation solving and convolutional neural networks (CNNs) for heat-flux reconstruction, improve computational efficiency but still only replace or accelerate individual components within the conventional thermal-analysis paradigm~\cite{aymerich2023physics,looby2019convolutional,wang2025revisiting}.
\textit{However, these methods remain dependent on thermal measurements, heat-conduction modeling, or simulation-derived data, limiting their direct application to online-oriented reconstruction.}
In contrast, macroscopic plasma-state signals are routinely recorded during plasma discharges and are usually available for online monitoring and control~\cite{kates2019predicting,guo2021disruption,zheng2023disruption,matos2020classification}.
This naturally raises an important question: \textit{Can time-resolved radial divertor heat-flux profiles be reconstructed directly from macroscopic plasma-state signals, without relying on conventional infrared-based inversion?}

As illustrated in Figure~\ref{fig:firstimg}(c), we introduce an online-oriented signal-based heat-flux reconstruction task that directly maps macroscopic plasma-state signals to time-resolved radial divertor heat-flux profiles.
To support this task, we construct \textbf{DivMPS2HF}, a multi-source discharge dataset that aligns heterogeneous plasma-state signals with radial heat-flux profiles.
We further propose \textbf{SafeDivertor}, which jointly models observed signals and target-channel placeholders.
Because the target-channel placeholders provide no heat-flux information, \textbf{physical prior-aware initialization (PPI)} supplies them with a channel-level statistical heat-flux prior, providing radial-distribution guidance before reconstruction.
Since heterogeneous plasma-state signals may differ in reliability and cause the model to over-rely on a small subset of channels, \textbf{input perturbation (IP)} regularizes the conditioning process by perturbing randomly selected observed channels during training.
Moreover, because point-wise reconstruction loss alone tends to smooth rapid heat-flux variations, \textbf{spectral-aware reconstruction optimization (SRO)} exploits the time-frequency prior embodied in the ground-truth heat-flux profiles through a multi-scale short-time Fourier transform (STFT) loss, thereby preserving transient dynamics and spectral consistency across multiple temporal scales.
As these objectives emphasize different aspects of reconstruction, \textbf{progressive training (PT)} introduces them in stages to stabilize optimization. 
Experiments on \textbf{DivMPS2HF} show that \textbf{SafeDivertor} achieves the best overall performance among the evaluated time-series baselines across all five metrics.
The main contributions of this work are summarized as follows:

\begin{itemize}
    \item To the best of our knowledge, we are the \textbf{first} to formulate an online-oriented signal-based reconstruction paradigm that shifts divertor heat-flux analysis from offline infrared-based inversion toward online-oriented reconstruction from macroscopic plasma-state signals available during discharge.

    \item We propose \textbf{SafeDivertor}, a task-driven framework that integrates physical prior-aware initialization, input perturbation, spectral-aware reconstruction optimization, and progressive training for faithful heat-flux reconstruction.

    \item We construct \textbf{DivMPS2HF}, a multi-source discharge dataset and evaluation benchmark for this task, and establish a strong reference baseline that achieves the best overall performance among the evaluated time-series methods across all five metrics.
\end{itemize}

\section{Related Work}

\textbf{Divertor Heat-Flux Analysis.~}  
Accurate divertor heat-flux analysis is essential for understanding plasma-wall interactions and protecting plasma-facing components in magnetic-confinement fusion devices.
Existing studies mainly rely on infrared thermal imaging technology.
First, the surface temperature of the divertor target plate is measured by an IR camera.
Then, the heat flux is inferred by solving the heat conduction equations related to material properties and boundary conditions~\cite{lee2017thermographic,herrmann1995energy}.
For example, KSTAR employs the NANTHELOT heat-flux reconstruction code to solve the heat diffusion equation from measured tile surface temperature, while EAST calculates divertor heat-flux distributions from IR measurements using DFLUX or FEM-based solvers~\cite{lee2017thermographic,shi2017heat}.
These pipelines are often computationally expensive and are commonly used for offline post-shot analysis or require additional acceleration for real-time deployment.
To improve efficiency, recent studies have introduced learning-based methods into heat-flux analysis. 
For example, approaches based on physics-informed neural networks (PINNs) have been explored to accelerate heat-equation solving toward real-time heat-flux estimation, while convolutional neural network (CNN)-based methods have been used to reconstruct parameterized heat-flux profiles from simulation-generated thermocouple signals~\cite{aymerich2023physics,looby2019convolutional}.
The method based on neural networks has also been applied to plasma diagnostic signals for cross-device disruption prediction~\cite{kates2019predicting}, EAST disruption prediction~\cite{guo2021disruption}, and four-class ELM recognition using HGTS-Former on EAST-ELM640~\cite{si2025hgts}. 
Deep neural networks have further predicted ELM onset from turbulence signals~\cite{joung2024tokamak}, while time-series extrinsic regression has reconstructed missing electron-temperature diagnostics~\cite{wang2025time}.
However, most existing heat-flux analysis methods either rely on thermal imaging measurements or focus on solving heat-conduction equations, while direct reconstruction of time-resolved divertor heat-flux profiles from multi-source discharge signals remains less explored.

\textbf{Multivariate Time-Series Modeling.~}  
Multivariate time series modeling has been extensively studied to capture temporal dependencies and cross-variable interactions across multiple channels~\cite{song2024deep,kim2025comprehensive,torres2021deep}.
In recent years, general time series models have achieved significant performance from different modeling perspectives.
For example, iTransformer treats each variate as a token by inverting the dimension to enhance cross-variable dependency modeling~\cite{liu2024itransformer}. 
Timer-XL~\cite{liu2025timer} represents time series as a long context token sequence and uses Transformer for multivariate sequence modeling.
SimpleTM~\cite{chen2025simpletm} combines signal processing ideas with a lightweight attention structure to provide a simple yet effective multivariate prediction baseline.
TimeFilter~\cite{hu2025timefilter} captures local dynamic patterns by constructing a spatiotemporal graph structure and performing patch-level filtering.
Frequency-domain and multi-periodic representations have also been explored
for general time-series modeling. 
FreTS~\cite{yi2023frequency} performs inter-series and intra-series
learning through frequency-domain MLPs to capture global temporal and
cross-variable dependencies. 
FiLM~\cite{zhou2022film} combines frequency
enhancement with Legendre projection to improve long-term time-series
forecasting. 
TimesNet~\cite{wu2023timesnet} transforms one-dimensional time
series into two-dimensional representations according to multiple periods,
thereby modeling both intra-period and inter-period variations.
TimePerceiver~\cite{lee2026timeperceiver} uses an encoder-decoder structure and learnable queries to adapt to prediction targets at different time locations. 
Sonnet~\cite{shu2026sonnet} captures temporal dynamics in the spectral domain and captures cross-variable relationships by utilizing spectral consistency between variables.
Some methods also focus on time series prediction under exogenous variable conditions. 
For example, TimeXer~\cite{wang2024timexer} fuses endogenous and exogenous variable information through patch-wise self-attention and variate-wise cross-attention, while GCGNet~\cite{li2026gcgnet} uses graph structure consistency constraints to model the correlation between exogenous and target variables.
Meanwhile, time-series imputation provides useful insights into recovering unavailable variables from observed contexts. 
Representative works such as T1~\cite{park2026t} combine a CNN-Transformer hybrid architecture with a channel-head binding mechanism to achieve robust imputation.
Although these methods provide useful paradigms for multivariate time-series modeling, divertor heat-flux reconstruction differs from standard forecasting, exogenous-variable prediction, and partial imputation, because all target heat-flux channels are unavailable throughout the current input window and must be reconstructed from physically different macroscopic plasma-state signals.

\section{Motivation}

Unlike conventional time-series imputation~\cite{du2023saits,cao2018brits}, which reconstructs partially missing observations within the same variable space, our task requires reconstructing entirely unavailable heat-flux channels from physically distinct macroscopic plasma-state signals. Therefore, we formulate it as a structured channel-level reconstruction problem, as follows.

Let $\mathcal{D}=\left\{\left(\mathbf{X}^{(i)},\mathbf{Y}^{(i)}\right)\right\}_{i=1}^{N}$ denote a discharge dataset, where $N$ denotes the number of samples after window sampling.
For the $i$-th sample, $\mathbf{X}^{(i)}$ represents the macroscopic plasma-state signals within a time window of length $T$, and $\mathbf{Y}^{(i)}$ denotes the corresponding time-resolved heat-flux profile on the divertor target plate. 
The input sequence is defined as $\mathbf{X}^{(i)}=[x_1^{(i)},x_2^{(i)},...,x_C^{(i)}] \in \mathbb{R}^{C \times T}$, where $C=67$ is the number of signal channels; 
and the corresponding target heat-flux profile is defined as $\mathbf{Y}^{(i)}=[y_1^{(i)},y_2^{(i)},...,y_O^{(i)}] \in \mathbb{R}^{O \times T}$, where $O=116$ is the number of heat-flux channels.

Since the actual heat-flux sequence within the current time window is unobservable, we further represent this task as a structured channel-level reconstruction problem based on target-channel initialization.
Specifically, an initial representation of the target heat-flux channels, denoted as $\mathbf{Y}^{(i)}_{\mathrm{init}} \in \mathbb{R}^{O \times T}$, is concatenated with the observed signals $\mathbf{X}^{(i)}$ along the channel dimension to form a unified input sequence:
\begin{equation}
\label{eq:init}
    \mathbf{Z}^{(i)}=[\mathbf{X}^{(i)},\mathbf{Y}^{(i)}_{\mathrm{init}}] \in \mathbb{R}^{F \times T}, 
    \quad
    F = C + O.
\end{equation}
It is important to note that $\mathbf{Y}_{\mathrm{init}}^{(i)}$ does not contain the actual heat-flux information of the current sample; it is only used as an initial placeholder for the target heat-flux channels. 
The model reconstructs the actual heat-flux profile based on this unified input sequence:
\begin{equation}
\label{eq:modeling}
    \hat{\mathbf{Y}}^{(i)}=f_{\boldsymbol{\theta}}(\mathbf{Z}^{(i)}), 
    \quad
    f_{\boldsymbol{\theta}}:\mathbb{R}^{F\times T}\rightarrow\mathbb{R}^{O\times T}.
\end{equation}
The model parameters are optimized by minimizing the reconstruction error between the reconstructed heat-flux sequence and the actual heat-flux sequence:
\begin{equation}
\label{eq:opti}
    \boldsymbol{\theta}^{*}=\arg\min_{\boldsymbol{\theta}}
    \frac{1}{N}
    \sum_{i=1}^{N}
    \mathcal{L}
    \left(\hat{\mathbf{Y}}^{(i)},\mathbf{Y}^{(i)}\right).
\end{equation}
During training and evaluation, the loss and all metrics are computed only on the $O$ target heat-flux channels.

Although the task formulation is straightforward, it presents several task-specific challenges.
First, the target-channel placeholders contain no heat-flux information, leaving the model without explicit guidance about the radial distribution to be reconstructed.
Second, the heterogeneous plasma-state signals differ in physical meaning and measurement reliability, which may cause the model to over-rely on a small subset of channels.
Third, point-wise reconstruction losses tend to smooth rapid heat-flux variations and fail to explicitly exploit the time-frequency prior contained in the target profiles.
Finally, these objectives emphasize different aspects of reconstruction and are difficult to optimize simultaneously. 
These challenges motivate \textbf{SafeDivertor}, which introduces \textbf{PPI}, \textbf{IP}, \textbf{SRO}, and \textbf{PT} to address the lack of target guidance, over-reliance on specific input channels, insufficient spectral preservation, and optimization instability, respectively, as detailed in the following section.

\begin{figure*}[!t]
    \centering
    \includegraphics[width=1.0\linewidth]{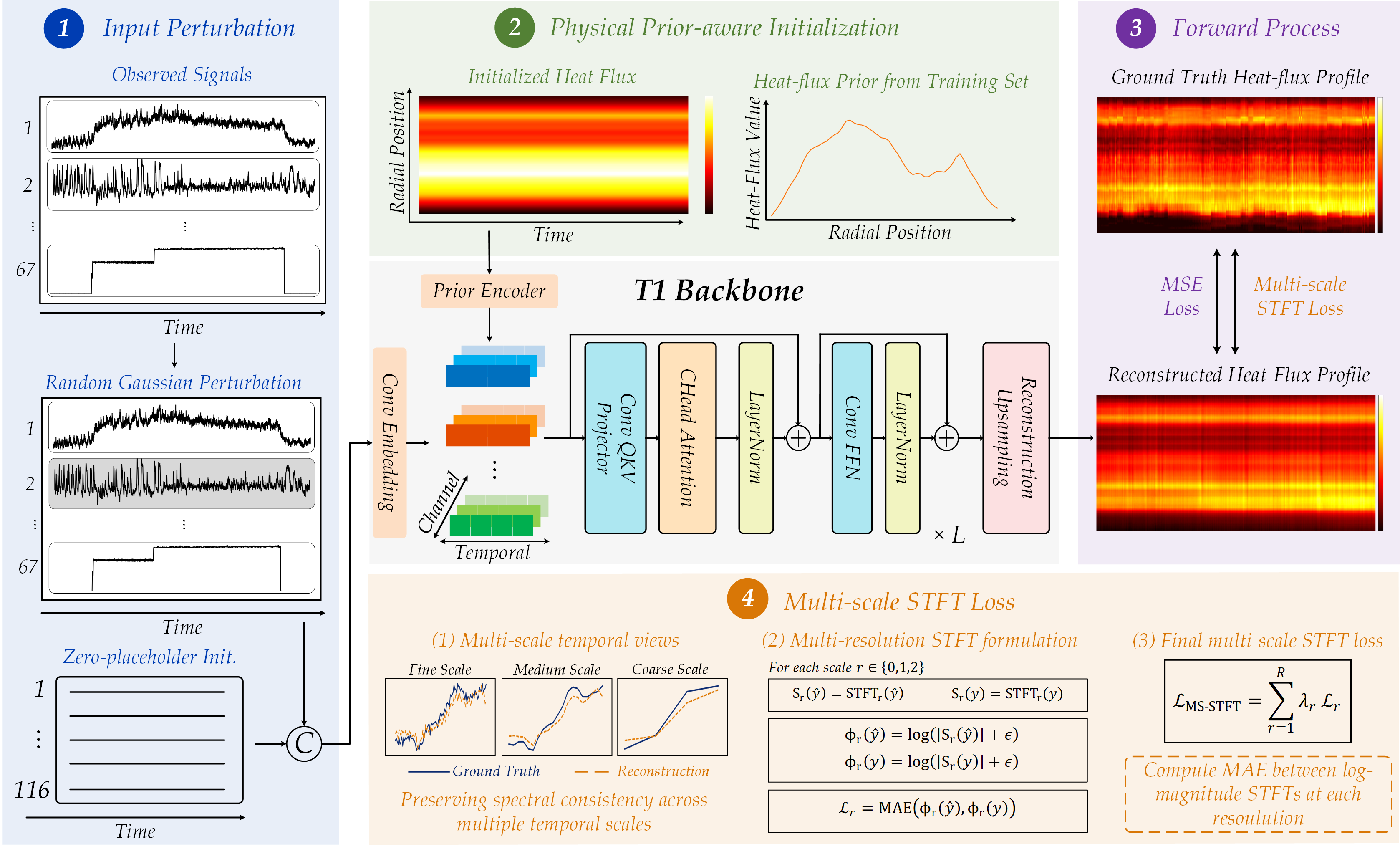}
    \caption{Overview of the proposed SafeDivertor framework for divertor heat-flux reconstruction from macroscopic plasma-state signals. SafeDivertor integrates physical prior-aware initialization, input perturbation, T1 backbone reconstruction, and multi-scale STFT-based spectral supervision.}
    \label{fig:method}
\end{figure*}

\section{Methodology}
\subsection{Dataset Curation}
To support the proposed online-oriented reconstruction task, we construct \textbf{DivMPS2HF}, a multi-source discharge dataset containing $77$ shots, divided into $64$, $3$, and $10$ shots for training, validation, and testing, respectively.
The split is performed at the shot level to prevent cross-shot information leakage. 
All signals are temporally aligned and resampled to a unified $1~\mathrm{kHz}$ grid.
Each sample contains $67$ input channels, including plasma-state diagnostic signals, equilibrium and configuration parameters, and signals related to external actuators and device control. 
These channels are selected from a broader pool of candidate signals to retain signals with reliable data quality, strong physical relevance to divertor heat-load evolution, and useful predictive information.
The reconstruction target is a time-resolved radial heat-flux profile consisting of $116$ channels, obtained from post-shot heat-flux analysis as supervised labels. 
Notably, infrared measurements and heat-conduction inversion results are used only to construct the reference labels and are not included in the model input. 
During inference, SafeDivertor directly reconstructs the heat-flux profile from macroscopic plasma-state signals through a single neural forward pass. 

\subsection{Physical Prior-aware Initialization}
In this paper, we model the multi-source signals and target heat-flux channels as a unified structured time series.
The first $67$ channels represent the observed multi-source signals, while the last $116$ channels are zero-valued placeholders for the target heat-flux channels at different radial locations on the divertor target plate.
However, target-channel placeholders contain no heat-flux distribution information and fail to reflect the inherent differences in basic heat loads at different radial locations.

Therefore, we introduce \textbf{PPI} and construct a channel-level heat-flux prior from the training set to provide statistical guidance for the unavailable target channels, as illustrated in Part $2$ of Figure~\ref{fig:method}.
Specifically, for the $o$-th target heat-flux channel, its statistical prior is computed as the average heat-flux value over all training shots and time steps:
\begin{equation}
\label{eq:mu_o}
\mu_o
=
\frac{1}{\sum_{s=1}^{S_{\mathrm{train}}}L_s}
\sum_{s=1}^{S_{\mathrm{train}}}
\sum_{t=1}^{L_s}
\mathbf{Y}^{(s)}_{o,t},
\quad
o=1,\ldots,O,
\end{equation}
where $S_{\mathrm{train}}$ denotes the number of training discharge shots, 
$L_s$ denotes the length of the $s$-th training shot, 
and $Y_{o,t}^{(s)}$ is the heat-flux value of the $o$-th target channel at time step $t$ in the $s$-th training shot.
For each sample, the radial heat-flux prior is defined as follows:
\begin{equation}
\label{eq:prior}
\boldsymbol{\mu}
=
[\mu_1,\mu_2,\ldots,\mu_O]^\top,
\quad
\mathbf{P}
=
\boldsymbol{\mu}\mathbf{1}_{1 \times T} \in \mathbb{R}^{O \times T},
\end{equation}
where $\mathbf{1}_{1 \times T}\in\mathbb{R}^{1 \times T}$ denotes an all-ones row vector and $T$ is the length of the input time window.

Based on this prior, we introduce a lightweight prior encoding branch to extract structural information.
Specifically, the original input with zero-valued target placeholders is first encoded into a feature representation $\mathbf{H}_0$.
According to the channel organization of the unified input, $\mathbf{H}_0$ is divided into the observed-signal features and target-placeholder features:
\begin{equation}
\label{eq:feature_split}
\mathbf{H}_0^{(i)}
=
\left[
\mathbf{H}_{0,\mathrm{obs}}^{(i)},
\mathbf{H}_{0,\mathrm{tar}}^{(i)}
\right].
\end{equation}
Meanwhile, the statistical heat-flux prior $\mathbf{P}$ is processed by a convolutional prior encoder $g_{\boldsymbol{\theta}}(\cdot)$ to obtain the prior feature.
The prior feature is then added only to the target-placeholder features, yielding the prior-enhanced representation:
\begin{equation}
\label{eq:representation}
\mathbf{H}^{(i)}
=
\left[
\mathbf{H}_{0,\mathrm{obs}}^{(i)},
\mathbf{H}_{0,\mathrm{tar}}^{(i)}
+
g_{\boldsymbol{\theta}}(\mathbf{P})
\right].
\end{equation}
Here, $\mathbf{H}^{(i)}$ denotes the prior-enhanced
feature representation. This design preserves the observed-signal
features and injects the statistical radial heat-flux information only
into the target-channel representations through an independent prior
encoding branch.

\subsection{Input Perturbation}
In actual discharge data acquisition, multi-source signals are inevitably affected by factors such as hardware noise, acquisition errors, and signal fluctuations. 
The observation quality of different channels may also change with the shot and time window. 
To reduce the model's over-reliance on specific input channels and improve its tolerance to channel-level noise, we introduce \textbf{IP} during the training phase.
As illustrated in Part 1 of Figure~\ref{fig:method}, Gaussian perturbation is applied only to selected observable signals, while the remaining observable signals and the heat-flux prior are kept unchanged. 
This design simulates noisy or unreliable measurements.
Specifically, for each training sample, we randomly select $20\%$ of the $67$ observable signals and add Gaussian noise with a standard deviation of $0.03$ to the complete time series of these channels. 
The augmented input is defined as follows:
\begin{equation}
\label{eq:channel_noise_aug}
\tilde{\mathbf{X}}_{c,t}
=
\begin{cases}
\mathbf{X}_{c,t}+\epsilon_{c,t}, 
& c \in \mathcal{C}_{n}, \\
\mathbf{X}_{c,t}, 
& c \notin \mathcal{C}_{n},
\end{cases}
\quad
\epsilon_{c,t}\sim\mathcal{N}(0,\sigma^2),
\end{equation}
where $\mathbf{\tilde{X}_{c,t}}$ and $\mathbf{X}_{c,t}$ denote the augmented and original signals of channel $c$ at time step $t$, $\mathcal{C}_n$ is the randomly selected subset containing $20\%$ of the channels,
and $\epsilon_{c,t}$ denotes the added Gaussian noise. 
$\sigma$ controls the noise intensity and is set to $0.03$ in our experiments.

\subsection{Spectral-aware Reconstruction Optimization}

\noindent \textbf{Basic Reconstruction Loss.} Given the reconstructed heat-flux profile $\hat{\mathbf{Y}}$ and the true heat-flux profile $\mathbf{Y}$, we first adopt the mean squared error as the basic reconstruction loss:
\begin{equation}
\label{eq:basic_loss}
    \mathcal{L}_\mathrm{MSE}=\frac{1}{BOT}\sum_{b=1}^B \sum_{o=1}^O \sum_{t=1}^T \left ( \hat{\mathbf{Y}}_{b,o,t}-\mathbf{Y}_{b,o,t}\right)^2,
\end{equation}
where $B$ is the batch size, $O=116$ denotes the number of target heat-flux channels, and $T$ is the length of the time window.
The MSE loss provides point-wise supervision for numerical fidelity in the time domain.

\noindent \textbf{Multi-scale STFT-based Spectral Loss.} 
Although the MSE loss provides direct point-wise supervision in the time domain,  it mainly measures numerical deviations at each time step and does not explicitly constrain the frequency characteristics of the reconstructed heat-flux sequences.
In divertor heat-load evolution, rapid variations and transient events may appear as high-frequency components in the heat-flux profile.
Models optimized solely by MSE may produce overly smoothed reconstructions that fail to preserve these spectral details.

Therefore, we introduce \textbf{SRO}, which employs a spectral loss based on the multi-scale short-time Fourier transform (STFT).
Compared to single-scale spectral constraints, multi-scale designs are better suited for heat-flux reconstruction because different time scales emphasize different dynamic modes.
Shorter temporal scales are more sensitive to rapid local variations and transient heat-load dynamics, while longer temporal scales help capture broader heat-flux evolution trends.
Therefore, applying spectral constraints at multiple temporal scales encourages the model to preserve both local high-frequency dynamics and global time-frequency consistency.
Let $\phi(\cdot)$ denote the log-magnitude STFT operator along the temporal dimension:
\begin{equation}
\label{eq:stft}
    \phi(\mathbf{Y})=\log \left (\left |\mathrm{STFT}(\mathbf{Y}) \right |+\epsilon \right ),
\end{equation}
where $\mathrm{STFT}(\cdot)$ denotes the STFT operation along the temporal dimension, and $\epsilon$ is a small constant for numerical stability. 
The multi-scale STFT loss is formulated as follows:
\begin{equation}
\label{eq:stft_loss}
    \mathcal{L}_\mathrm{MS-STFT} = \sum_{r=1}^R \lambda_r\mathrm{MAE}\left(\phi(\mathbf{Y}_r),\phi(\hat{\mathbf{Y}}_r)\right),
\end{equation}
where $R=3$, $\mathbf{Y}_r$ and $\hat{\mathbf{Y}}_r$ 
denote the ground-truth and reconstructed heat-flux sequences at the $r$-th temporal scale, respectively, and $\lambda_r$ is the weight for the $r$-th temporal scale. 
By matching the log-magnitude spectra of heat-flux sequences at multiple temporal scales, this loss helps the model preserve both rapid high-frequency variations and broader temporal spectral patterns.

\noindent \textbf{Overall Training Objective.}
The final training objective combines the basic reconstruction loss and the multi-scale STFT-based spectral loss. 
The MSE loss provides direct supervision for numerical accuracy in the time domain, while the multi-scale STFT loss constrains the spectral characteristics of heat-flux sequences across different temporal scales. 
The overall loss is defined as follows:
\begin{equation}
\label{eq:overal_loss}
    \mathcal{L}_\mathrm{total} = \mathcal{L}_\mathrm{MSE} + \mathcal{L}_\mathrm{MS-STFT}.
\end{equation}
By jointly optimizing these objectives, the model improves the point-wise numerical accuracy and multi-scale spectral consistency of the reconstructed heat-flux profile.

\subsection{Progressive Training}
Directly applying input perturbation and spectral-aware reconstruction optimization from the beginning may increase the optimization difficulty, since the model needs to simultaneously learn the basic reconstruction mapping, robustness to input perturbations, and spectral consistency.
To stabilize the training process, we adopt a three-stage training strategy. 
In the first stage, the model is trained only with the basic MSE reconstruction loss, allowing it to learn a stable mapping from multi-source signals to heat-flux profiles.
In the second stage, channel-level noise augmentation is enabled while the model continues to optimize the reconstruction objective, which improves its robustness to input-channel perturbations.
In the third stage, the input perturbation is disabled, and the multi-scale STFT loss is introduced to refine the reconstructed heat-flux profiles under clean input conditions. 
This stage focuses on improving spectral consistency and preserving rapid heat-flux variations.
The model parameters are progressively inherited across stages, enabling stable optimization from basic numerical reconstruction to robust and spectral-aware heat-flux reconstruction.

\section{Experiments}

\subsection{Dataset, Metric, and Baselines}

\textbf{Dataset.~} We construct \textbf{DivMPS2HF}, a multi-source time-series dataset for reconstructing divertor heat-flux profiles from macroscopic plasma-state signals. The dataset contains data from $77$ discharge shots, among which $64$ shots are used for training, $3$ shots for validation, and $10$ shots for testing. To avoid cross-shot information leakage, the dataset is split at the discharge-shot level, ensuring that samples from the same discharge shot do not appear in different subsets. 

Before constructing the dataset, all selected signals are aligned to a unified temporal grid with a sampling rate of $1~\mathrm{kHz}$. 
The selected signals have heterogeneous sampling rates: most signals are recorded around $1~\mathrm{kHz}$, while some diagnostic signals are sampled at either lower or much higher frequencies. For example, $\mathcal{D}_\alpha$-related signals can be sampled at up to $250~\mathrm{kHz}$. 
To obtain synchronized multi-source inputs, we resample all signals to the common $1~\mathrm{kHz}$ temporal grid before model training and evaluation.
Signals with higher sampling rates are downsampled, while signals with lower sampling rates are mapped to the aligned time grid using temporal resampling.

After temporal alignment, samples are generated on the fly during data loading by applying a sliding window of $T=500$ time steps to each discharge shot, corresponding to $0.5~\mathrm{s}$ at $1~\mathrm{kHz}$. 
A stride of one time step is used to densely capture the temporal evolution of the heat-flux profiles. 
This sampling procedure yields approximately $150{,}504$, $11{,}838$, and $28{,}982$ windows for training, validation, and testing, respectively.

The model input consists of 67 channels of macroscopic plasma-state signals, which include not only traditional diagnostic measurements but also plasma state diagnostic signals, equilibrium and configuration parameters and signals related to external actuators and device control.
These signals characterize the global and boundary plasma states related to the evolution of divertor heat load during the discharge process.

The reconstruction target is the \textbf{time-resolved heat-flux profile} on the divertor target. For each time step, the heat-flux profile is represented as a 116-dimensional vector, corresponding to 116 sampling channels in the radial direction of the target. The reference heat-flux profiles are obtained from post-shot heat-flux analysis and are used as supervised labels for model training and evaluation. It should be emphasized that infrared images and heat-conduction inversion results are not used as model inputs. During inference, SafeDivertor only takes macroscopic plasma-state signals as input and reconstructs the target heat-flux profile through a direct neural forward pass. The DivMPS2HF dataset will be made available upon reasonable request for academic research purposes.

\textbf{Metrics.~} 
We evaluate the reconstructed divertor heat-flux profiles using five metrics: Mean Squared Error (\textbf{MSE}), Mean Absolute Error (\textbf{MAE}),  Structural Similarity (\textbf{SSIM})~\cite{wang2004image}, Log-Spectral Distance (\textbf{LSD})~\cite{li2026audio}, and Log-Spectral Distance of High Frequency (\textbf{LSD-HF}).
\textbf{MSE} and \textbf{MAE} are used as standard point-wise error metrics to evaluate the point-wise numerical accuracy of reconstructed heat flux in the time domain. 
Given $N$ test samples, the reconstructed heat-flux profiles $\mathbf{\hat{Y}} \in \mathbb{R}^{N \times O \times T}$ and the ground-truth heat-flux profiles $\mathbf{Y} \in \mathbb{R}^{N \times O \times T}$, \textbf{MSE} and \textbf{MAE} are defined as follows:
\begin{equation}
    \label{eq:metric_mse}
    \mathrm{MSE}=\frac{1}{NOT}\sum_{n=1}^{N}\sum_{o=1}^{O}\sum_{t=1}^{T}\left (\mathbf{Y}_{n,o,t}-\mathbf{\hat{Y}}_{n,o,t} \right )^2,
\end{equation}
\begin{equation}
    \label{eq:metric_mae}
    \mathrm{MAE}=\frac{1}{NOT}\sum_{n=1}^{N}\sum_{o=1}^{O}\sum_{t=1}^{T}\left |\mathbf{Y}_{n,o,t}-\mathbf{\hat{Y}}_{n,o,t} \right |.
\end{equation}
Here, $O=116$ denotes the number of target heat-flux channels, and $T$ denotes the length of the time window.

\textbf{SSIM} is used to evaluate the structural similarity of reconstructed spatiotemporal heat flux maps, taking into account both temporal evolution and radial profile distribution.
For each sample, the heat-flux profile is treated as a two-dimensional map over radial channels and time steps. 
We first compute SSIM on each sample and then average the scores over all test samples:
\begin{equation}
    \label{eq:ssim_score}
    \mathrm{SSIM}
    =
    \frac{1}{N}
    \sum_{n=1}^{N}
    \mathrm{SSIM}
    \left(
    \hat{\mathbf{Y}}_n,
    \mathbf{Y}_n
    \right),
\end{equation}
where $N$ denotes the number of test samples. 
For the $n$-th sample, the sample-level SSIM is computed by averaging the local SSIM values over all evaluated $7\times7$ blocks:
\begin{equation}
\label{eq:ssim_metric}
\mathrm{SSIM}
\left(
\hat{\mathbf{Y}}_n,
\mathbf{Y}_n
\right)
=
\frac{1}{K}
\sum_{k=1}^{K}
\mathrm{SSIM}_{n,k},
\end{equation}
where $K$ is the number of evaluated local blocks in each heat-flux map. 
The local SSIM of the $k$-th block is defined as
\begin{equation}
\label{eq:ssim_local}
\mathrm{SSIM}_{n,k}
=
\frac{
(2\mu_{\mathbf{Y}_{n,k}}\mu_{\hat{\mathbf{Y}}_{n,k}}+\epsilon_1)
(2\mathrm{Cov}(\mathbf{Y}_{n,k},\hat{\mathbf{Y}}_{n,k})+\epsilon_2)
}{
(\mu_{\mathbf{Y}_{n,k}}^{2}+\mu_{\hat{\mathbf{Y}}_{n,k}}^{2}+\epsilon_1)
(\sigma_{\mathbf{Y}_{n,k}}^{2}+\sigma_{\hat{\mathbf{Y}}_{n,k}}^{2}+\epsilon_2)
}.
\end{equation}
Here, $\mathbf{Y}_{n,k}$ and $\hat{\mathbf{Y}}_{n,k}$ denote the $k$-th local $7\times7$ block of the ground-truth and reconstructed heat-flux maps for the $n$-th test sample, respectively. 
$\mu$, $\sigma^2$, and $\mathrm{Cov}(\cdot,\cdot)$ denote the local mean, variance, and covariance, and $\epsilon_1$, $\epsilon_2 $ are small constants for numerical stability.

Let $\mathbf{S}$ and $\hat{\mathbf{S}}$ denote the power spectra of the ground-truth and reconstructed heat-flux profiles after STFT:
\begin{equation}
\label{eq:power_spectrum}
\mathbf{S}=|\mathrm{STFT}(\mathbf{Y})|^2+\epsilon,
\quad
\hat{\mathbf{S}}=|\mathrm{STFT}(\hat{\mathbf{Y}})|^2+\epsilon,
\end{equation}
where $\epsilon$ is a small constant for numerical stability.

For each evaluated spectrum, we first compute the log-spectral distance:
\begin{equation}
\label{eq:local_lsd}
d_{n,o,m}
=
\sqrt{
\frac{1}{F}
\sum_{f=1}^{F}
\left(
\log \hat{\mathbf{S}}_{n,o,f,m}
-
\log \mathbf{S}_{n,o,f,m}
\right)^2
},
\end{equation}
where $F$ denotes the number of frequency bins. 
The final \textbf{LSD} score is obtained by averaging over all test samples, target channels, and STFT frames:
\begin{equation}
\label{eq:lsd}
\mathrm{LSD}
=
\frac{1}{NOM}
\sum_{n=1}^{N}
\sum_{o=1}^{O}
\sum_{m=1}^{M}
d_{n,o,m},
\end{equation}
where $N$ is the number of test samples, $O$ is the number of target heat-flux channels, and $M$ is the number of STFT frames.

Similarly, \textbf{LSD-HF} is computed by restricting the frequency bins to the high-frequency set $\mathcal{F}_{\mathrm{HF}}$:
\begin{equation}
\label{eq:local_lsd_hf}
d^{\mathrm{HF}}_{n,o,m}
=
\sqrt{
\frac{1}{|\mathcal{F}_{\mathrm{HF}}|}
\sum_{f\in\mathcal{F}_{\mathrm{HF}}}
\left(
\log \hat{\mathbf{S}}_{n,o,f,m}
-
\log \mathbf{S}_{n,o,f,m}
\right)^2
},
\end{equation}
\begin{equation}
\label{eq:lsd_hf}
\mathrm{LSD\text{-}HF}
=
\frac{1}{NOM}
\sum_{n=1}^{N}
\sum_{o=1}^{O}
\sum_{m=1}^{M}
d^{\mathrm{HF}}_{n,o,m}.
\end{equation}
All metrics are computed only on the 116 target heat-flux channels.

\textbf{Baselines.~}
We select representative state-of-the-art time series models from recent years as benchmarks for comparison.
Specifically, we include general multivariate time series forecasting models, including Sonnet~\cite{shu2026sonnet}, Timer-XL~\cite{liu2025timer}, TimeFilter~\cite{hu2025timefilter}, SimpleTM~\cite{chen2025simpletm}, TimePerceiver~\cite{lee2026timeperceiver}, and iTransformer~\cite{liu2024itransformer}.
Furthermore, since our task naturally involves using multi-source signals as external conditions to reconstruct the target heat flux sequence, we also include exogenous variable prediction models, including CrossLinear~\cite{zhou2025crosslinear} and TimeXer~\cite{wang2024timexer}.
Additionally, we use T1~\cite{park2026t} as the time series imputation benchmark, where multi-source signal channels are treated as observed variables, and heat flux channels are treated as structured missing variables.

\subsection{Implementation Details}
We construct \textbf{DivMPS2HF}, a multi-source time-series dataset for reconstructing divertor heat-flux profiles from macroscopic plasma-state signals. 
All models are evaluated using five metrics: \textbf{MSE}, \textbf{MAE}, \textbf{SSIM}, \textbf{LSD}, \textbf{LSD-HF}, which measure point-wise numerical accuracy, structural similarity, and spectral consistency of the reconstructed heat-flux profiles. 

During training and evaluation, fixed-length windows are sampled from the corresponding discharge shots. The input window length is set to $T=500$ time steps, corresponding to a $0.5$-second discharge segment under the unified $1~\mathrm{kHz}$ sampling rate. The experiments are conducted on eight NVIDIA RTX $4090$ GPUs. For SafeDivertor, the Gaussian perturbation ratio is set to $20\%$, and the noise standard deviation is set to $0.03$. The channel-level heat-flux prior is computed only from the training set and kept fixed during validation and testing.  
For efficiency evaluation, all models are tested with batch size $1$ and input length $500$, and the reported inference latency only measures the model forward process.

\subsection{Main Results}

As shown in Table~\ref{tab:main_results}, SafeDivertor achieves the best performance across all evaluation metrics compared with representative time-series baselines.
Among the baselines, T1~\cite{park2026t} shows the strongest overall performance in numerical accuracy and spectral consistency, achieving the best baseline MSE, MAE, LSD, and LSD-HF, while TimeFilter~\cite{hu2025timefilter} obtains the highest baseline SSIM.
\begin{table} 
\caption{Overall comparison with representative time-series baselines on the DivMPS2HF dataset. 
}
\label{tab:main_results}
\resizebox{\linewidth}{!}{
\begin{tabular}{l|ccccc}
\toprule
\textbf{Models}        
& \textbf{MSE}$\downarrow$  
& \textbf{MAE}$\downarrow$   
& \textbf{SSIM}$\uparrow$  
& \textbf{LSD}$\downarrow$   
& \textbf{LSD-HF}$\downarrow$ 
\\ \midrule
\textbf{T1}            & \underline{0.262}    & \underline{0.333}    & 0.848          & \underline{3.521}    & \underline{3.829}     
\\ \midrule
\textbf{Sonnet}        & 0.382          & 0.414          & 0.804          & 7.294          & 7.418           
\\ \midrule
\textbf{Timer-XL}      & 0.307          & 0.367          & 0.824          & 4.718          & 5.420           
\\ \midrule
\textbf{TimeFilter}    & 0.279          & 0.354          & \underline{0.854}    & 3.868          & 4.573           
\\ \midrule
\textbf{CrossLinear}   & 0.378          & 0.427          & 0.706          & 5.101          & 4.976           
\\ \midrule
\textbf{SimpleTM}      & 0.540          & 0.499          & 0.760          & 8.033          & 8.153           
\\ \midrule
\textbf{TimePerceiver} & 0.437          & 0.426          & 0.814          & 4.546          & 5.359           
\\  \midrule
\textbf{TimeXer}       & 0.538          & 0.488          & 0.769          & 4.876          & 5.228           
\\ \midrule
\textbf{iTransformer}  & 0.469          & 0.446          & 0.792          & 7.893          & 8.136           
\\ \midrule
\textbf{SafeDivertor}  & \textbf{0.200} & \textbf{0.291} & \textbf{0.870} & \textbf{2.475} & \textbf{2.729} 
 \\ \bottomrule
\end{tabular}
}
\end{table}
In contrast, several forecasting and exogenous-variable forecasting models show limited performance, indicating that directly applying existing time-series forecasting paradigms is insufficient for reconstructing target heat-flux channels from heterogeneous macroscopic plasma-state signals.

Compared with the best baseline result for each metric, SafeDivertor reduces MSE from $0.262$ to $0.200$, MAE from $0.333$ to $0.291$, LSD from $3.521$ to $2.475$, and LSD-HF from $3.829$ to $2.729$, while improving SSIM from $0.854$ to $0.870$. 
These results demonstrate that the proposed task-driven designs help SafeDivertor better preserve numerical accuracy, spatiotemporal structure, and spectral characteristics of divertor heat-flux profiles.

\subsection{Ablation Study}
Table~\ref{tab:ablation} presents the ablation results of the proposed task-driven components, including physical prior-aware initialization (\textbf{PPI}), input perturbation (\textbf{IP}), spectral-aware reconstruction optimization (\textbf{SRO}), and progressive training (\textbf{PT}). 
The baseline model without these components achieves an MSE of $0.262$, MAE of $0.333$, SSIM of $0.848$, LSD of $3.521$, and LSD-HF of $3.829$.
\begin{table}[!t]
\centering
\caption{Ablation study of the main components in SafeDivertor on the DivMPS2HF dataset.}
\label{tab:ablation}
\resizebox{\linewidth}{!}{
\begin{tabular}{c|cccc|ccccc}
\toprule
&\textbf{PPI} 
&\textbf{IP}
&\textbf{SRO}
&\textbf{PT}
&\textbf{MSE}$\downarrow$  
&\textbf{MAE}$\downarrow$ 
&\textbf{SSIM}$\uparrow$  
&\textbf{LSD}$\downarrow$  
&\textbf{LSD-HF}$\downarrow$
\\ \midrule
\textbf{\#1} & \xmark  & \xmark  & \xmark & \xmark   &0.262   &0.333   &0.848    &3.521    &3.829      
\\  
\textbf{\#2} & \cmark  & \xmark & \xmark & \xmark  &\underline{0.211}   &\underline{0.293}   &\textbf{0.870}    &3.617    &4.141       
\\ 
\textbf{\#3} & \xmark  & \cmark & \xmark & \xmark &0.230  &0.313   &0.860    &3.725    &4.205
\\ 
\textbf{\#4} & \xmark  & \xmark  & \cmark & \xmark  &0.219    &0.299  &0.863    &\textbf{2.306}    &\textbf{2.511} 
\\

\textbf{\#5} & \cmark  & \cmark  & \cmark & \xmark  &0.211 &0.296 &\underline{0.869} &\underline{2.326} &\underline{2.568} 
\\ 
\textbf{\#6} & \cmark  & \cmark  & \cmark  & \cmark &\textbf{0.200} &\textbf{0.291} &\textbf{0.870} &2.475 &2.729  \\
\bottomrule
\end{tabular}
}
\end{table} 
When each component is introduced individually, different performance trends can be observed. 
\textbf{PPI} reduces MSE from $0.262$ to $0.211$, MAE from $0.333$ to $0.293$, and improves SSIM from $0.848$ to $0.870$, indicating that the physical heat-flux prior provides useful statistical radial-distribution information for unavailable target channels. 
\textbf{IP} also improves the time-domain and structural metrics, reducing MSE to $0.230$ and improving SSIM to $0.860$, which shows that perturbing observable signal channels helps regularize the model's dependence on specific signals. 
However, both \textbf{PPI} and \textbf{IP} lead to higher LSD and LSD-HF, because they do not explicitly supervise the spectral characteristics of reconstructed heat-flux sequences. 
In contrast, \textbf{SRO} contributes most significantly to spectral consistency, reducing LSD from $3.521$ to $2.306$ and LSD-HF from $3.829$ to $2.511$. 
This confirms the necessity of multi-scale STFT-based spectral supervision for preserving spectral characteristics and high-frequency transient variations.

When \textbf{PPI}, \textbf{IP}, and \textbf{SRO} are jointly used, the model achieves a more balanced performance across time-domain, structural, and spectral metrics.
This demonstrates that these components play complementary roles: \textbf{PPI} provides statistical heat-flux prior information,
\textbf{IP} regularizes heterogeneous observable signals, and \textbf{SRO} enhances spectral consistency.
After further introducing \textbf{PT}, the full model achieves the best MSE and MAE of $0.200$ and $0.291$, respectively, while maintaining competitive spectral performance. 
These results show that the proposed task-driven components improve complementary aspects of reconstruction quality.

\begin{figure*}[]
    \centering
    \includegraphics[width=1.0\linewidth]{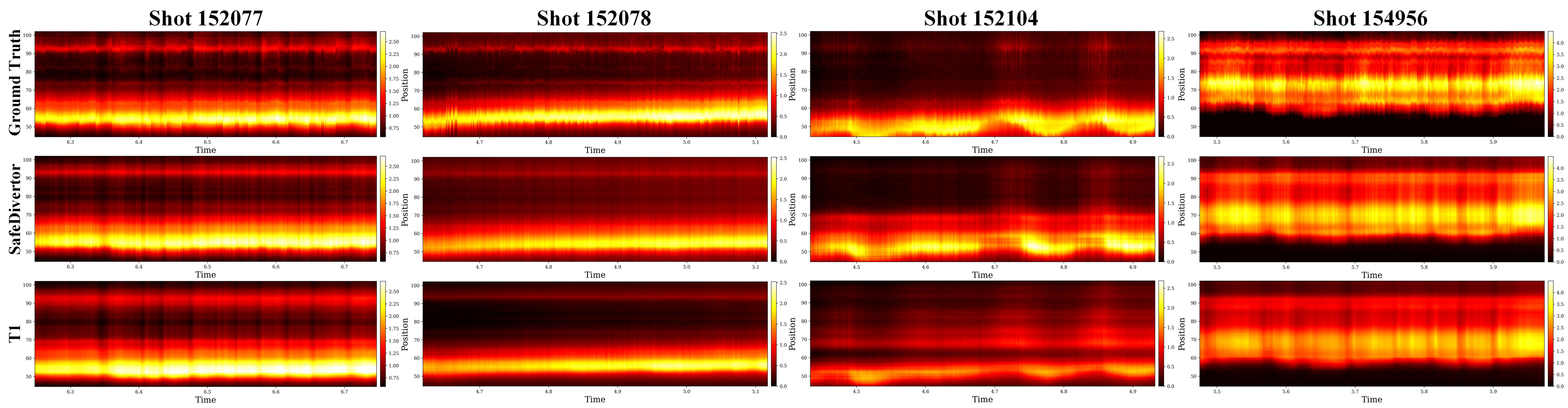}
    \caption{Qualitative visualization of four randomly selected samples from different discharge shots, showing the ground truth, SafeDivertor, and T1 from top to bottom.}
    \label{fig:visualization}
\end{figure*}

\subsection{Effect of Multi-scale Spectral Loss}
As shown in Table ~\ref{tab:loss_ablation}, we analyze the role of spectral loss in the spectral-aware reconstruction optimization module.
The \textbf{\textit{w/o} SRO} setting represents a model trained using only the basic MSE reconstruction loss.
Introducing single-scale STFT loss significantly improves the model across all metrics.
This indicates that the spectral optimization objective helps the model better preserve the spectral characteristics and rapid changes in the heat-flux sequences.
Compared to the single-scale setting, the multi-scale STFT loss function further improves most metrics.
These results demonstrate that applying spectral constraints across multiple timescales is more effective than using a single scale, as different scales can capture both broader heat flux evolution patterns and local high-frequency variations.
Overall, the multi-scale STFT loss function provides a better balance between numerical accuracy and spectral consistency.

\begin{table}
\centering
\caption{Ablation study of the multi-scale STFT loss.}
\label{tab:loss_ablation}
\resizebox{\linewidth}{!}{
\begin{tabular}{l|ccc} 
\toprule
Metrics    &\textbf{\textit{w/o} SRO}       & \textbf{Single Scale} & \textbf{Multi Scale} \\
\midrule
\textbf{MSE}$\downarrow$    & 0.262 & \underline{0.221}  & \textbf{0.219}      \\
\textbf{MAE}$\downarrow$     & 0.333 & \underline{0.301}  & \textbf{0.299}      \\
\textbf{SSIM}$\uparrow$   & 0.848 & \textbf{0.867}  & \underline{0.863}      \\
\textbf{LSD}$\downarrow$     & 3.521 & \underline{2.367}  & \textbf{2.306}      \\
\textbf{LSD-HF}$\downarrow$  & 3.829 & \underline{2.549}  & \textbf{2.511}      \\
\bottomrule
\end{tabular}
}
\end{table}

\subsection{Input Perturbation Strategy Analysis}

\begin{table}
\centering
\caption{Effect of different input perturbation strategies. CM and GP denote channel masking and Gaussian perturbation, respectively.}
\label{tab:strategies}
\resizebox{\linewidth}{!}{
\begin{tabular}{l|c|cc|cc|cc}
\toprule
&\textbf{\textit{w/o} IP}  
& \multicolumn{2}{c|}{\textbf{r=0.1}} 
& \multicolumn{2}{c|}{\textbf{r=0.2}}
& \multicolumn{2}{c}{\textbf{r=0.4}} \\
\midrule
\textbf{Metrics} 
&\text{-}& \textbf{CM}      
& \textbf{GP}     
& \textbf{CM}      
& \textbf{GP}     
& \textbf{CM}      
& \textbf{GP}     \\
\midrule

\textbf{MSE}    &0.262    &0.251  & \underline{0.229}   
& 0.288     & 0.230   & 0.276     & \textbf{0.203}  
\\

\textbf{MAE}   &0.333  & 0.329   & 0.316    & 0.345   & \underline{0.313}     & 0.339    & \textbf{0.297}  \\

\textbf{SSIM}  &0.848  & 0.858            & \underline{0.861}     & 0.854            & 0.860           & 0.853            & \textbf{0.874}  \\

\textbf{LSD}   &\textbf{3.521}  &3.653      & 3.816           & 3.696            & 3.725           & \underline{3.599}   & 3.784           \\

\textbf{LSD-HF} &\textbf{3.829} &4.093      & 4.315           & 4.096            & 4.205           & \underline{4.013}   & 4.163     \\
\bottomrule
\end{tabular}
}
\end{table}
We further evaluate different input perturbation strategies by applying channel masking and Gaussian perturbation to observable signal channels under different perturbation ratios $ r \in \{0.1,0.2,0.4\}$. 
As shown in Table~\ref{tab:strategies}, channel masking provides limited and unstable improvements. 
Although it slightly improves MSE and SSIM at $r=0.1$, larger masking ratios degrade the numerical metrics, suggesting that directly removing observable channels may discard useful plasma-state information.
In contrast, Gaussian perturbation consistently improves numerical accuracy and structural similarity across different ratios, indicating that perturbing observable channels is more effective than completely masking them. Although both strategies degrade spectral performance, this is expected because they do not explicitly constrain frequency-domain characteristics.
Overall, Gaussian perturbation is a more effective input perturbation strategy for improving numerical accuracy and structural similarity.

\begin{figure*}[!hbtp]
\centering
\includegraphics[width=1.0\linewidth]{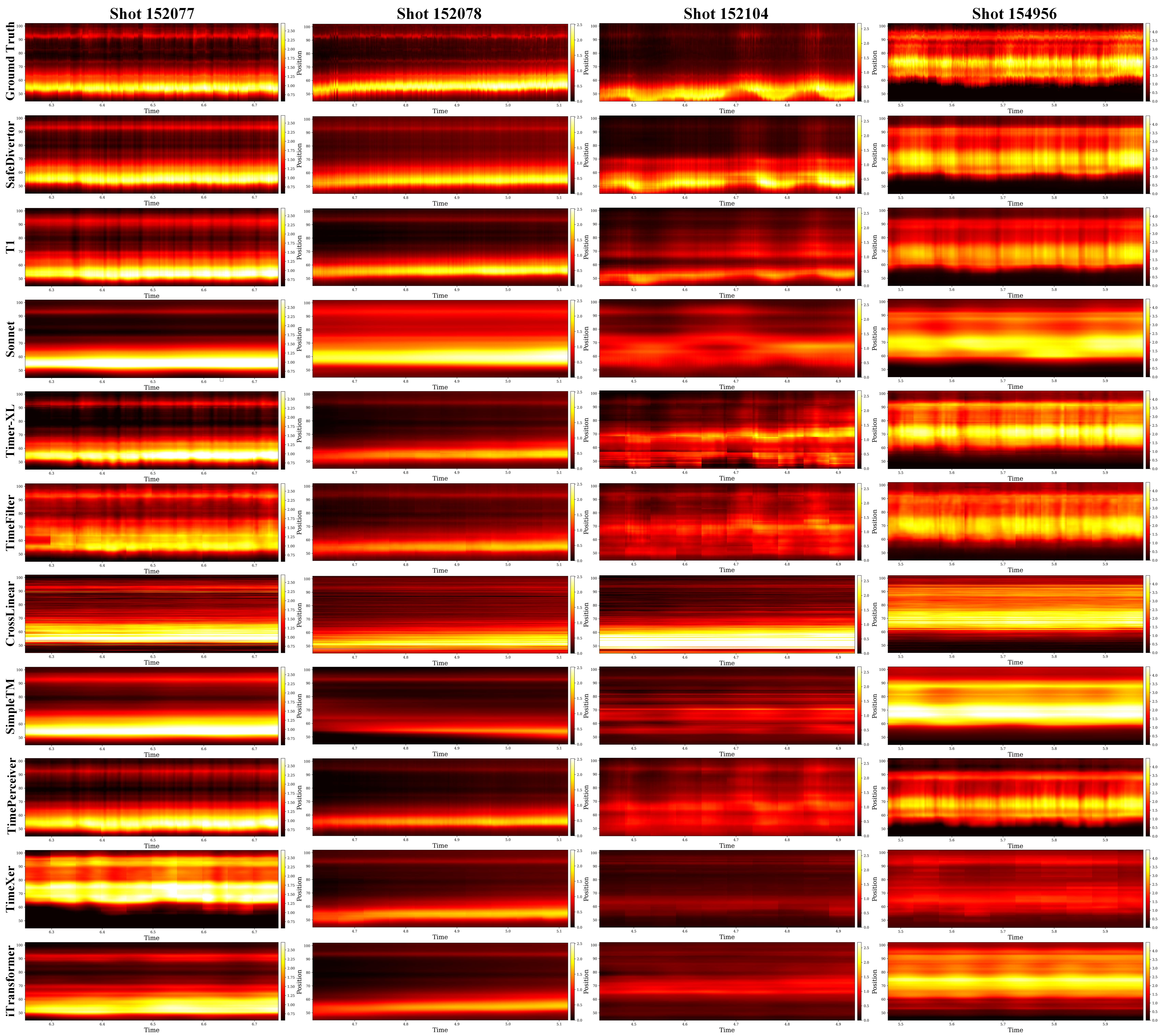}
\caption{Qualitative comparison of four samples, one from each of four test shots. Each column corresponds to one sample, while the rows show the ground truth and reconstruction results of SafeDivertor and the evaluated baselines.}
\label{fig:vis_all}
\end{figure*}

\subsection{Efficiency Analysis}
Table~\ref{tab:efficiency} compares the computational complexity and inference efficiency of different models. 
SafeDivertor requires $21.255$ GFLOPs and achieves an inference latency of $11.313$ ms per sample with a batch size of $1$ and an input length of $500$. 
Since each input window corresponds to a $0.5$-second discharge segment at $1~\mathrm{kHz}$, this latency indicates its potential for online-oriented window-level heat-flux reconstruction. 
Compared with the T1 backbone, SafeDivertor retains the same parameter scale, with GFLOPs increasing only from $21.240$ to $21.255$ and comparable latency, while substantially improving reconstruction performance. 
These results demonstrate a favorable performance-efficiency trade-off with negligible additional inference cost.
\begin{table}[]
\centering
\caption{Comparison of model parameters (M), computational cost (GFLOPs), and per-sample inference latency (ms).}
\label{tab:efficiency}
\resizebox{\linewidth}{!}{
\begin{tabular}{l|lll} 
\toprule
Model         & \textbf{Params.}    & \textbf{GFLOPs}   & \textbf{Inference} \\
\midrule
Sonnet        & 42.239 & 123.188 & 10.947    \\
T1            & \underline{12.783} & 21.240   & 11.023    \\
Timer-XL      & 33.825 & 61.583  & 14.721    \\
TimeFilter    & \textbf{11.011} & 35.118  & 261.343   \\
CrossLinear   & 33.130  & \underline{5.122}   & 4.623     \\
SimpleTM      & 34.753 & 21.629  & 6.126     \\
TimePerceiver & 16.032 & 14.824  & \underline{2.530}      \\
TimeXer       & 51.088 & 44.684  & 4.161     \\
iTransformer  & 34.635 & \textbf{2.390}    & \textbf{1.732} \\          
SafeDivertor & \underline{12.783} & 21.255  & 11.313    \\
\bottomrule
\end{tabular}
}
\end{table}

\subsection{Visualization}
As shown in Figure~\ref{fig:visualization}, we compare the ground truth, SafeDivertor, and T1 on four samples.
The original T1 backbone recovers the overall spatiotemporal heat-flux structure and main radial high-heat-flux bands, but its results are relatively over-smoothed and may underestimate local high-intensity regions, failing to preserve temporal fluctuations and fine-grained heat-flux variations.
In contrast, SafeDivertor better preserves the main heat-flux structures while recovering clearer high-intensity bands and richer temporal details.
These visual results are consistent with the quantitative improvements in Table~\ref{tab:main_results}, further demonstrating the advantage of the proposed task-driven designs. 

Figure~\ref{fig:vis_all} presents qualitative comparisons on four samples from different test shots. Most baselines can identify the approximate radial locations of the dominant heat-flux bands, but their reconstruction characteristics vary considerably. Several methods produce over-smoothed profiles with limited temporal variation, while others exhibit noticeable intensity bias, discontinuous patterns, or horizontal artifacts. Although the original T1 backbone preserves the overall spatiotemporal structure relatively well, it still tends to underestimate local high-intensity regions and suppress fine-grained temporal fluctuations. In contrast, SafeDivertor more faithfully recovers the positions and intensities of the main heat-flux bands while retaining clearer transient variations across different discharge conditions. These qualitative results are consistent with the quantitative improvements reported in the main experiments, further demonstrating the effectiveness of the proposed task-driven designs.

\section{Conclusion}

In this paper, we introduce a new online-oriented paradigm for directly reconstructing divertor heat-flux profiles from macroscopic plasma-state signals, without using infrared surface-temperature measurements as model inputs. 
We also construct \textbf{DivMPS2HF}, a multi-source discharge dataset for this task, and propose \textbf{SafeDivertor}, a unified reconstruction framework equipped with physical prior-aware initialization, input perturbation, spectral-aware reconstruction optimization, and progressive training. 
Experiments on DivMPS2HF show that SafeDivertor achieves the best overall performance among the evaluated time-series baselines across all five metrics. Efficiency analysis further shows that SafeDivertor maintains inference efficiency comparable to T1, providing a favorable latency basis for future online deployment. Future work will expand the dataset, improve cross-scenario generalization, and further enhance reconstruction efficiency.

\bibliographystyle{IEEEtran}
\bibliography{ref}

\begin{thebibliography}{10}
\providecommand{\url}[1]{#1}
\csname url@samestyle\endcsname
\providecommand{\newblock}{\relax}
\providecommand{\bibinfo}[2]{#2}
\providecommand{\BIBentrySTDinterwordspacing}{\spaceskip=0pt\relax}
\providecommand{\BIBentryALTinterwordstretchfactor}{4}
\providecommand{\BIBentryALTinterwordspacing}{\spaceskip=\fontdimen2\font plus
\BIBentryALTinterwordstretchfactor\fontdimen3\font minus
  \fontdimen4\font\relax}
\providecommand{\BIBforeignlanguage}[2]{{%
\expandafter\ifx\csname l@#1\endcsname\relax
\typeout{** WARNING: IEEEtran.bst: No hyphenation pattern has been}%
\typeout{** loaded for the language `#1'. Using the pattern for}%
\typeout{** the default language instead.}%
\else
\language=\csname l@#1\endcsname
\fi
#2}}
\providecommand{\BIBdecl}{\relax}
\BIBdecl

\bibitem{yang2020development}
Z.~Yang, P.~He, H.~Yan, Z.~Bin, S.~Shuangbao, F.~Wang, M.~Chen, G.~Jia, and
  X.~Gong, ``The development of a three-dimensional finite element method code
  for the heat flux analysis of tungsten monoblock divertor on east,''
  \emph{Fusion Engineering and Design}, vol. 152, p. 111448, 2020.

\bibitem{pitts2019physics}
R.~A. Pitts, X.~Bonnin, F.~Escourbiac, H.~Frerichs, J.~Gunn, T.~Hirai,
  A.~Kukushkin, E.~Kaveeva, M.~Miller, D.~Moulton \emph{et~al.}, ``Physics
  basis for the first iter tungsten divertor,'' \emph{Nuclear Materials and
  Energy}, vol.~20, p. 100696, 2019.

\bibitem{aymerich2023physics}
E.~Aymerich, F.~Pisano, B.~Cannas, G.~Sias, A.~Fanni, Y.~Gao,
  D.~B{\"o}ckenhoff, M.~Jakubowski \emph{et~al.}, ``Physics informed neural
  networks towards the real-time calculation of heat fluxes at w7-x,''
  \emph{Nuclear Materials and Energy}, vol.~34, p. 101401, 2023.

\bibitem{herrmann1995energy}
A.~Herrmann, W.~Junker, K.~Gunther, S.~Bosch, M.~Kaufmann, J.~Neuhauser,
  G.~Pautasso, T.~Richter, and R.~Schneider, ``Energy flux to the asdex-upgrade
  diverter plates determined by thermography and calorimetry,'' \emph{Plasma
  Physics and Controlled Fusion}, vol.~37, no.~1, pp. 17--29, 1995.

\bibitem{lee2017thermographic}
H.~Lee, R.~Pitts, C.~Kang, S.~Oh, J.~Bak, S.~Hong, H.~Wi, Y.~Kim, H.~Kim,
  D.~Seo \emph{et~al.}, ``Thermographic studies of outer target heat fluxes on
  kstar,'' \emph{Nuclear Materials and Energy}, vol.~12, pp. 541--547, 2017.

\bibitem{shi2017heat}
B.~Shi, Z.-D. Yang, B.~Zhang, C.~Yang, K.-F. Gan, M.-W. Chen, J.-H. Yang,
  H.~Zhang, J.-L. Qi, X.-Z. Gong \emph{et~al.}, ``Heat flux on east divertor
  plate in h-mode with lhcd/lhcd+ nbi,'' \emph{Chinese Physics Letters},
  vol.~34, no.~9, p. 095201, 2017.

\bibitem{looby2019convolutional}
T.~Looby, M.~Reinke, D.~Donovan, T.~Gray, M.~Messineo, and A.~Khodak,
  ``Convolutional neural networks for heat flux model validation on nstx-u,''
  \emph{IEEE Transactions on Plasma Science}, vol.~48, no.~1, pp. 3--13, 2019.

\bibitem{wang2025revisiting}
X.~Wang, Z.~Yan, H.~Si, Z.~Yang, Q.~Yang, D.~Sun, W.~Lyu, and J.~Tang,
  ``Revisiting heat flux analysis of tungsten monoblock divertor on east using
  physics-informed neural network,'' \emph{arXiv preprint arXiv:2508.03776},
  2025.

\bibitem{kates2019predicting}
J.~Kates-Harbeck, A.~Svyatkovskiy, and W.~Tang, ``Predicting disruptive
  instabilities in controlled fusion plasmas through deep learning,''
  \emph{Nature}, vol. 568, no. 7753, pp. 526--531, 2019.

\bibitem{guo2021disruption}
B.~Guo, B.~Shen, D.~Chen, C.~Rea, R.~Granetz, Y.~Huang, L.~Zeng, H.~Zhang,
  J.~Qian, Y.~Sun \emph{et~al.}, ``Disruption prediction using a full
  convolutional neural network on east,'' \emph{Plasma Physics and Controlled
  Fusion}, vol.~63, no.~2, p. 025008, 2021.

\bibitem{zheng2023disruption}
W.~Zheng, F.~Xue, Z.~Chen, D.~Chen, B.~Guo, C.~Shen, X.~Ai, N.~Wang, M.~Zhang,
  Y.~Ding \emph{et~al.}, ``Disruption prediction for future tokamaks using
  parameter-based transfer learning,'' \emph{Communications Physics}, vol.~6,
  no.~1, p. 181, 2023.

\bibitem{matos2020classification}
F.~Matos, V.~Menkovski, F.~Felici, A.~Pau, F.~Jenko, T.~Team, and E.~M. Team,
  ``Classification of tokamak plasma confinement states with convolutional
  recurrent neural networks,'' \emph{Nuclear Fusion}, vol.~60, no.~3, p.
  036022, 2020.

\bibitem{si2025hgts}
H.~Si, X.~Wang, F.~Zhang, X.~Zhou, D.~Sun, W.~Lyu, Q.~Yang, and J.~Tang,
  ``Hgts-former: Hierarchical hypergraph transformer for multivariate time
  series analysis,'' \emph{arXiv preprint arXiv:2508.02411}, 2025.

\bibitem{joung2024tokamak}
S.~Joung, D.~R. Smith, G.~McKee, Z.~Yan, K.~Gill, J.~Zimmerman, B.~Geiger,
  R.~Coffee, F.~O’Shea, A.~Jalalvand \emph{et~al.}, ``Tokamak edge localized
  mode onset prediction with deep neural network and pedestal turbulence,''
  \emph{Nuclear Fusion}, vol.~64, no.~6, p. 066038, 2024.

\bibitem{wang2025time}
M.~Wang, C.~Wan, J.~Lu, Z.~Yu, B.~Xiao, Y.~Li, X.~He, Z.~Luo, Q.~Yuan, Y.~Hu
  \emph{et~al.}, ``Time series extrinsic regression for reconstructing missing
  electron temperature in tokamak,'' \emph{Nuclear Fusion}, vol.~65, no.~7, p.
  076008, 2025.

\bibitem{song2024deep}
X.~Song, L.~Deng, H.~Wang, Y.~Zhang, Y.~He, and W.~Cao, ``Deep learning-based
  time series forecasting: X. song et al.'' \emph{Artificial Intelligence
  Review}, vol.~58, no.~1, p.~23, 2024.

\bibitem{kim2025comprehensive}
J.~Kim, H.~Kim, H.~Kim, D.~Lee, and S.~Yoon, ``A comprehensive survey of deep
  learning for time series forecasting: architectural diversity and open
  challenges,'' \emph{Artificial Intelligence Review}, vol.~58, no.~7, p. 216,
  2025.

\bibitem{torres2021deep}
J.~F. Torres, D.~Hadjout, A.~Sebaa, F.~Mart{\'\i}nez-{\'A}lvarez, and
  A.~Troncoso, ``Deep learning for time series forecasting: a survey,''
  \emph{Big data}, vol.~9, no.~1, pp. 3--21, 2021.

\bibitem{liu2024itransformer}
Y.~Liu, T.~Hu, H.~Zhang, H.~Wu, S.~Wang, L.~Ma, and M.~Long, ``itransformer:
  Inverted transformers are effective for time series forecasting,'' in
  \emph{International conference on learning representations}, vol. 2024, 2024,
  pp. 11\,116--11\,140.

\bibitem{liu2025timer}
Y.~Liu, G.~Qin, X.~Huang, J.~Wang, and M.~Long, ``Timer-xl: Long-context
  transformers for unified time series forecasting,'' in \emph{International
  Conference on Learning Representations}, vol. 2025, 2025, pp.
  83\,982--84\,006.

\bibitem{chen2025simpletm}
H.~Chen, V.~Luong, L.~Mukherjee, and V.~Singh, ``Simple{TM}: A simple baseline
  for multivariate time series forecasting,'' in \emph{The Thirteenth
  International Conference on Learning Representations}, 2025.

\bibitem{hu2025timefilter}
Y.~Hu, G.~Zhang, P.~Liu, D.~Lan, N.~Li, D.~Cheng, T.~Dai, S.-T. Xia, and
  S.~Pan, ``Timefilter: Patch-specific spatial-temporal graph filtration for
  time series forecasting,'' in \emph{Forty-second International Conference on
  Machine Learning}, 2025.

\bibitem{yi2023frequency}
K.~Yi, Q.~Zhang, W.~Fan, S.~Wang, P.~Wang, H.~He, N.~An, D.~Lian, L.~Cao, and
  Z.~Niu, ``Frequency-domain mlps are more effective learners in time series
  forecasting,'' \emph{Advances in neural information processing systems},
  vol.~36, pp. 76\,656--76\,679, 2023.

\bibitem{zhou2022film}
T.~Zhou, Z.~Ma, Q.~Wen, L.~Sun, T.~Yao, W.~Yin, R.~Jin \emph{et~al.}, ``Film:
  Frequency improved legendre memory model for long-term time series
  forecasting,'' \emph{Advances in neural information processing systems},
  vol.~35, pp. 12\,677--12\,690, 2022.

\bibitem{wu2023timesnet}
H.~Wu, T.~Hu, Y.~Liu, H.~Zhou, J.~Wang, and M.~Long, ``Timesnet: Temporal
  2d-variation modeling for general time series analysis,'' in \emph{The
  Eleventh International Conference on Learning Representations}, 2023.

\bibitem{lee2026timeperceiver}
J.~Lee and H.~Lee, ``Timeperceiver: An encoder-decoder framework for
  generalized time-series forecasting,'' \emph{Advances in Neural Information
  Processing Systems}, vol.~38, pp. 136\,136--136\,165, 2026.

\bibitem{shu2026sonnet}
Y.~Shu and V.~Lampos, ``Sonnet: Spectral operator neural network for
  multivariable time series forecasting,'' in \emph{Proceedings of the AAAI
  Conference on Artificial Intelligence}, vol.~40, no.~30, 2026, pp.
  25\,419--25\,427.

\bibitem{wang2024timexer}
Y.~Wang, H.~Wu, J.~Dong, G.~Qin, H.~Zhang, Y.~Liu, Y.~Qiu, J.~Wang, and
  M.~Long, ``Timexer: Empowering transformers for time series forecasting with
  exogenous variables,'' \emph{Advances in Neural Information Processing
  Systems}, vol.~37, pp. 469--498, 2024.

\bibitem{li2026gcgnet}
Z.~Li, X.~Qiu, Y.~Zhu, X.~Wu, J.~Hu, C.~Guo, and B.~Yang, ``{GCGN}et:
  Graph-consistent generative network for time series forecasting with
  exogenous variables,'' in \emph{The Fourteenth International Conference on
  Learning Representations}, 2026.

\bibitem{park2026t}
D.~Park, H.~Ryu, S.~Bae, K.~Park, and H.-S. Kim, ``T1: One-to-one channel-head
  binding for multivariate time-series imputation,'' in \emph{The Fourteenth
  International Conference on Learning Representations}, 2026.

\bibitem{du2023saits}
W.~Du, D.~C{\^o}t{\'e}, and Y.~Liu, ``Saits: Self-attention-based imputation
  for time series,'' \emph{Expert Systems with Applications}, vol. 219, p.
  119619, 2023.

\bibitem{cao2018brits}
W.~Cao, D.~Wang, J.~Li, H.~Zhou, L.~Li, and Y.~Li, ``Brits: Bidirectional
  recurrent imputation for time series,'' \emph{Advances in neural information
  processing systems}, vol.~31, 2018.

\bibitem{wang2004image}
Z.~Wang, A.~C. Bovik, H.~R. Sheikh, and E.~P. Simoncelli, ``Image quality
  assessment: from error visibility to structural similarity,'' \emph{IEEE
  transactions on image processing}, vol.~13, no.~4, pp. 600--612, 2004.

\bibitem{li2026audio}
C.~Li, Z.~Chen, L.~Wang, and J.~Zhu, ``Audio super-resolution with latent
  bridge models,'' \emph{Advances in Neural Information Processing Systems},
  vol.~38, pp. 48\,593--48\,636, 2026.

\bibitem{zhou2025crosslinear}
P.~Zhou, Y.~Liu, J.~Liang, Q.~Song, and X.~Li, ``Crosslinear: Plug-and-play
  cross-correlation embedding for time series forecasting with exogenous
  variables,'' in \emph{Proceedings of the 31st ACM SIGKDD Conference on
  Knowledge Discovery and Data Mining V. 2}, 2025, pp. 4120--4131.

\end{thebibliography}

\end{document}